\documentclass{article}
\usepackage{spconf,amsmath,graphicx,pgfplots,tikz}
\usepackage{xfrac,bm,amssymb,multirow,algorithm,algorithmic}
\usepackage[caption=false]{subfig}
\pgfplotsset{compat=1.14}

\DeclareMathOperator*{\argmax}{arg\,max}
\DeclareMathOperator{\Tr}{Tr}

\newcommand{\bargraphrsme}[3] {

    \begin{tikzpicture}
        \tikzstyle{every node}=[font=\tiny]
        \begin{axis}[
            bar width=3pt,
            height=#3,
            symbolic y coords={MM,FFT01,FFT02,FFT04,FFT08,FFT16,FFT32,FFT01-QI,FFT02-QI,FFT04-QI,FFT08-QI,FFT16-QI,FFT32-QI,SVD},
            width=#2,
            xbar,
            xlabel={RMSE},
            xmin=0,
            xmax=0.35,
            ytick=data,
            y dir=reverse
            ]
            \addplot[blue,fill=blue] table {#1};
        \end{axis}
    \end{tikzpicture}    

}

\newcommand{\bargraphcpu}[3] {

    \begin{tikzpicture}
        \tikzstyle{every node}=[font=\tiny]
        \begin{axis}[
            bar width=3pt,
            height=#3,
            symbolic y coords={MM,FFT01,FFT02,FFT04,FFT08,FFT16,FFT32,FFT01-QI,FFT02-QI,FFT04-QI,FFT08-QI,FFT16-QI,FFT32-QI,SVD},
            width=#2,
            xbar,
            xlabel={Execution time per frame ($\mu$sec)},
            xmin=0,
            xmax=60,
            ytick=data,
            y dir=reverse
            ]
            \addplot[red,fill=red] table {#1};
        \end{axis}
    \end{tikzpicture}    

}

\title{A Study of the Complexity and Accuracy of Direction of Arrival Estimation Methods based on GCC-PHAT for a Pair of Close Microphones}
\name{Fran\c{c}ois Grondin, James Glass\thanks{This work was supported in part by the Toyota Research Institute and by the Fonds de recherche du Qu\'{e}bec –- Nature et technologies.}}
\address{Computer Science and Artificial Intelligence Laboratory\\
Massachusetts Institute of Technology\\
    Cambridge, MA 02139, USA \\
    \small\texttt{\{fgrondin,glass\}@mit.edu}}
\begin{document}
%\ninept
%
\maketitle
\begin{abstract}
This paper investigates the accuracy of various Generalized Cross-Correlation with Phase Transform (GCC-PHAT) methods for a close pair of microphones.
We investigate interpolation-based methods and also propose another approach based on Singular Value Decomposition (SVD).
All investigated methods are implemented in C code, and the execution time is measured to determine which approach is the most appealing for real-time applications on low-cost embedded hardware.
\end{abstract}

\section{Introduction}
\label{sec:intro}

Sound Source Localization (SSL) consists of estimating the Direction of Arrival (DOA) of a target source in space with respect to a microphone array.
DOA information is useful for audio scene analysis \cite{bregman1994auditory} and is often used by common beamforming methods \cite{habets2010new,johnson1993array,parra2002geometric,valin2004enhanced,yamamoto2005enhanced} to enhance the target source and reduce interferences.
One of the most popular SSL method relies on the Steered-Response Power Phase Transform (SRP-PHAT) \cite{brandstein1997robust,cobos2011modified}.
SRP-PHAT consists of computing the Generalized Cross-Correlation with Phase Transform (GCC-PHAT) on each pair of microphones \cite{kwon2010analysis}.
The Fast Fourier Transform (FFT) provides an efficient way to compute GCC-PHAT \cite{grondin2013manyears,valin2007robust}, which is central for real-time systems that rely on SRP-PHAT as the number of GCC-PHATs needed is proportional to the square of the number of microphones.
Using FFTs reduces the amount of computation, but also decreases the accuracy as the Time Difference of Arrival (TDOA) is rounded to the closest sample integer value.
Recent smart speakers such as the Amazon Echo, Google Home and Apple HomePod are equipped with multiple microphones spaced by a few centimeters \cite{agarwal2018opportunistic}.
With these devices, DOA estimation can be particulary challenging as microphones are close to each other in space and the sample rate is low.

To cope with the discretization artifacts, interpolation can be performed on the GCC-PHAT results \cite{jacovitti1993discrete,mccormick2013approach,viola2005spline,van2012time}.
Another approach consists of using fractional delay estimation \cite{maskell1999estimation}.
Finally, the fractional Fourier transform \cite{sharma2007time} attempts to overcome the FFT discretization drawback.

In this paper, we compare the accuracy of some interpolation-based methods with the exact GCC-PHAT computation that involves a significant amount of computation.
We propose another approach to estimate the DOA with two microphones based on Singular Value Decomposition (SVD).
All investigated methods are implemented in C code, and the executation time is also measured to determine which approach is the most appealing for real-time applications on low-cost embedded hardware.

\section{DOA Estimation Methods}
\label{sec:methods}

The direction of arrival (DOA) of a sound source can be derived directly from the Time Difference of Arrival (TDOA) between two microphones.
Let $f_S \in \mathbb{N}$ be the sample rate (in sample/sec), $d \in \mathbb{R}_{>0}$ the distance between two microphones, and $c \in \mathbb{R}_{>0}$ the speed of sound .
The TDOA $\tau \in \mathbb{R}$ (in sample) is given by (\ref{eq:methods_tau}), where the angle in radian $\theta \in \mathbb{R} \cap [-\pi/2, \pi/2]$ represents the DOA.
\begin{equation}
\tau = (f_S/c)d\sin\theta
\label{eq:methods_tau}
\end{equation}

The range $[-\pi/2, \pi/2]$ is discretized into Q points, which leads to discrete angles $\theta_q \in \mathbb{R} \cap [-\pi/2,\pi/2]$, where $q \in \mathbb{Z} \cap [0,Q-1]$:
\begin{equation}
\theta_q = \left(q/(Q-1)-1/2\right)\pi
\label{eq:methods_thetaq}
\end{equation}

The Short-Time Fourier Transform (STFT) provides a useful representation of the signals needed to compute the GCC-PHAT.
In this paper, the expression $N \in \mathbb{N}$ stands for the STFT frame size in samples and $\Delta N \in \mathbb{N}$ for the hop size in samples between successive frames.
We define $X_{m}^{l}[k]$ as the STFT coefficients of microphone $m \in \mathbb{N} \cap [1,2]$ for the frequency bins $k \in \mathbb{Z} \cap [0,N/2]$ at frame $l \in \mathbb{N}$.
Without loss of generality, we omit the frame index $l$ from now on for clarity.
The expression $X_{12}[k] \in \mathbb{C}$ then stands for the phase transformed cross-correlation spectrum at frequency bin $k$ between microphones $1$ and $2$:
\begin{equation}
X_{12}[k] = X_1[k]X_2[k]^*/(|X_1[k]||X_2[k]|)
\label{eq:methods_X12}
\end{equation}
where $\{\dots\}^*$ stands for the complex conjugate and $|\dots|$ for the absolute value.

Equation \ref{eq:methods_x12} computes the GCC-PHAT result, denoted as $x_{12}[q] \in \mathbb{R}$, where $j = \sqrt{-1}$ and $\mathcal{R}\{\dots\}$ extracts the real part only.
\begin{equation}
x_{12}[q] = \Re\left\{\sum_{k=0}^{N/2}{g_k X_{12}[k]\exp\left(j2\pi k \tau_q/N\right)}\right\}
\label{eq:methods_x12}
\end{equation}
where the scalar $g_k \in \mathbb{R}_{>0}$ defined in (\ref{eq:methods_g}) normalizes the transform.
\begin{equation}
g_k = \begin{cases}
1/\sqrt{N} & k = \{0, N/2\} \\
\sqrt{2/N} & 0 < k < N/2
\label{eq:methods_g}
\end{cases}
\end{equation}

The estimated DOA $\theta_{est}$ then corresponds to the discrete angle $\theta_{q_{max}}$ which $q$ maximizes the GCC-PHAT result, as shown in Eq.~\ref{eq:methods_qmax}.
\begin{equation}
q_{max} = \argmax_q{\{x_{12}[q]\}}
\label{eq:methods_qmax}
\end{equation}

It is also relevant to compute the maximum GCC-PHAT magnitude, denoted here as $E_{est} = x_{12}[q_{max}]$, which provides insightful information regarding true DOA and false detections.

\subsection{Matrix multiplication}

The most straightforward method consists in implementing Eq.~\ref{eq:methods_x12} via matrix multiplication, as shown in Eq.~\ref{eq:mm_x12}, where $\mathbf{x}_{12} \in \mathbb{R}^{Q \times 1}$, $\mathbf{X}_{12} \in \mathbb{C}^{(N/2+1) \times 1}$ and $\mathbf{W} \in \mathbb{C}^{Q \times (N/2+1)}$.
\begin{equation}
\mathbf{x}_{12} = \Re\left\{\mathbf{W}\mathbf{X}_{12}\right\}
\label{eq:mm_x12}
\end{equation}
Each element $w_{q,k} \in \mathbb{C}$ of $\mathbf{W}$ is computed offline, and given by Eq.~\ref{eq:mm_g}.
Note that the scaling factor ensures that $\textrm{diag}\{\mathbf{W}\mathbf{W}^H\} = \mathbf{I}_{Q}$, where the operator $\textrm{diag}\{\dots\}$ sets all non-diagonal elements to zero, $\{\dots\}^H$ stands for the Hermitian operator and $\mathbf{I}_{Q}$ is a $Q\times Q$ identity matrix.
\begin{equation}
w_{q,k} = g_k\exp\left(j2\pi k\tau_q/N\right)
\label{eq:mm_g}
\end{equation}

Although appealing in its simplicity and exactitude, this method involves a significant amount of computations, as the complexity order reaches $\mathcal{O}(QN)$.

\subsection{Fast Fourier Transform}

As an alternative to matrix multiplication, the inverse fast Fourier transform (IFFT) offers an approximation to Eq.~\ref{eq:methods_x12} where the TDOA $\tau_q$ is rounded to the closest integer ($\lfloor\tau_q\rceil \in \mathbb{Z}$).
To reduce the discretization error, it is possible to increase the spectrum size from $(N/2+1)$ to $(Ni/2+1)$ frequency bins, where the bins in the range $[N/2+2,Ni/2+1]$ are set to zero and $i$ is usually a power of $2$.
This zero padding in the frequency domain reduces the discretization error as it results in an interpolation operation after performing the IFFT, denoted by $\mathbf{y}_{12} \in \mathbb{R}^{N \times 1}$.
The value that corresponds to the closest TDOA is then mapped to each point $q$, as shown in Eq's.~\ref{eq:fft_y12}-\ref{eq:fft_map}.
\begin{equation}
\mathbf{y}_{12} = \textrm{IFFT}\{\mathbf{X}_{12}\}
\label{eq:fft_y12}
\end{equation}
\begin{equation}
\mathbf{x}_{12} = \textrm{map}\{\mathbf{y}_{12}\}
\label{eq:fft_x12}
\end{equation}
\begin{equation}
\textrm{map}: x_{12}[q] \leftarrow y_{12}[\lfloor i\tau_q\rceil \bmod iN]
\label{eq:fft_map}
\end{equation}

Note that the modulo operator in Eq.~\ref{eq:fft_map} ensures that negative values are mapped within the range $[0,iN-1]$.
Most general purpose processors (GPDs) and digital signal processors (DSPs) compute efficiently radiix-2 IFFTs, and therefore this approach leads a complexity order of $\mathcal{O}(iN\log iN)$.

\subsection{Quadratic interpolation}

A complementary approach to the IFFT consists of performing quadratic interpolation on signal $\mathbf{y}_{12}$ obtained in Eq.~\ref{eq:fft_y12}.
The rounded TDOA corresponds to $\hat{\tau}_q = \lfloor i\tau_q \rceil$, and the rounding error is defined as $\Delta\hat{\tau}_q \in \mathbb{R}$, where $\Delta\hat{\tau}_q = i\tau_q - \lfloor i\tau_q \rceil$.  For each point $q$, a parabolic curve is fit to the $\mathbf{y}_{12}$ samples associated to $\hat{\tau}_q$ and its neighbors.
The expression in Eq.~\ref{eq:qi_abc} then estimates the parabolic curve parameters $a_q \in \mathbb{R}$, $b_q \in \mathbb{R}$, and $c_q \in \mathbb{R}$.
\begin{equation}
\begin{bmatrix}
a_q \\ b_q \\ c_q
\end{bmatrix} = \frac{1}{2}
\begin{bmatrix}
1 & -2 & 1 \\
-1 & 0 & 1 \\
0 & 2 & 0 \\
\end{bmatrix}
\begin{bmatrix}
y[ (\hat{\tau}_q - 1) \bmod iN] \\ y[\hat{\tau}_q \bmod iN] \\ y[(\hat{\tau}_q + 1) \bmod iN]
\end{bmatrix}
\label{eq:qi_abc}
\end{equation}

Finally, the parabolic expression provides an estimation of the fractional TDOA from the rounding error $\Delta\hat{\tau}_q$:
\begin{equation}
x_{12}[q] = a_q(\Delta\hat{\tau}_q)^2 + b_q(\Delta\hat{\tau}_q) + c_q
\end{equation}

The complexity of computations is similar to the complexity of the IFFT-based method, with an additional parabolic fitting step for all $Q$ points, which results in a total complexity of $\mathcal{O}(iN\log iN + Q)$.

\subsection{Singular Value Decomposition}

Another alternative to matrix multiplication consists of decomposing the matrix $\mathbf{W}$ into a different set of bases.
As the microphone distance gets smaller, the bases that compose this matrix become more linearly dependent.
This implies that a smaller set of orthogonal bases obtained via Singular Value Decomposition (SVD) can span the same space.
The matrix $\mathbf{W}$ is first represented by the sum of two real matrices $\mathbf{W}_R \in \mathbb{R}^{Q \times (N/2+1)}$ and $\mathbf{W}_I \in \mathbb{R}^{Q \times (N/2+1)}$, such that $\mathbf{W} = \mathbf{W}_{R} + j \mathbf{W}_{I}$.
This step is optional but allows using existing SVD algorithms designed for real matrices only.
Given $\alpha \in \{R,I\}$, the decomposition of $\mathbf{W}_{\alpha}$ with the $K_{\alpha}$ largest singular values leads to the approximation in Eq.~\ref{eq:svd_Walpha}.
\begin{equation}
\mathbf{W}_{\alpha} \approx \mathbf{U}_{\alpha}\mathbf{S}_{\alpha}\mathbf{V}^T_{\alpha} = \mathbf{U}_{\alpha} \mathbf{T}_{\alpha}
\label{eq:svd_Walpha}
\end{equation}
where $\mathbf{U}_{\alpha} \in \mathbb{R}^{Q \times K_{\alpha}}$, $\mathbf{T}_{\alpha} \in \mathbb{R}^{K_{\alpha} \times N/2+1}$ and $\{\dots\}^{T}$ stands for the matrix transpose operator.

To reduce the amount of computations, it is desirable to make the rank $K_{\alpha}$ as small as possible, yet it needs to be large enough to ensure an accurate reconstruction of $\mathbf{W}_{\alpha}$.
This can be achieved by selecting the smallest $K_{\alpha}$ that satisfies the condition in Eq.~\ref{eq:svd_trace}, where the operator $\textrm{Tr}\{\dots\}$ provides the trace of the matrix, and the parameter $\delta$ is a small positive value that models the tolerable reconstruction error.
\begin{equation}
\Tr{\{\mathbf{S}_{\alpha}\mathbf{S}_{\alpha}^T\}} \geq (1 - \delta)\Tr{\{\mathbf{W}_{\alpha}\mathbf{W}_{\alpha}^T\}}
\label{eq:svd_trace}
\end{equation}

The decomposition obtained in Eq.~\ref{eq:svd_Walpha} is then substituted in the initial matrix multiplication in Eq.~\ref{eq:mm_x12}, which leads to Eq.~\ref{eq:svd_x12}, where $\Re\{\dots\}$ and $\Im\{\dots\}$ capture respectively the real and imaginary parts of the corresponding expression.
\begin{equation}
\mathbf{x}_{12} \approx \mathbf{U}_{R} \left(\mathbf{T}_{R} \Re\left\{\mathbf{X}_{12}\right\}\right) - \mathbf{U}_{I} \left(\mathbf{T}_{I} \Im\left\{\mathbf{X}_{12}\right\}\right)
\label{eq:svd_x12}
\end{equation}

The matrices $\mathbf{U}_R$, $\mathbf{T}_R$, $\mathbf{U}_R$ and $\mathbf{U}_I$ are computed offline using SVD, and then the expression in Eq.~\ref{eq:svd_x12} is evaluated online.
This leads to a complexity of $\mathcal{O}((K_R + K_I)(N + Q))$, which represents a significant reduction compared to matrix multiplication when $(K_R + K_I) \ll (N + Q)$.

Table \ref{tab:svd_summary} presents a summary of all methods and their corresponding complexity order.
It is interesting to note that the number of potential source positions $Q$ affects the complexity order of Matrix Multiplication, Quadratic Interpolation and Singular Value Decomposition, as opposed to the Fast Fourier Transform method.
Although no arithmetic is involved, there are $Q$ lookups during the mapping procedure described in (\ref{eq:fft_map}) for the FFT-based approach.

\begin{table}[!ht]
    \caption{Complexity Orders of DOA Estimation Methods}
    \vspace{2pt}
    \centering
    \def\arraystretch{1.0}
    \begin{tabular}{|c|c|}
        \hline
        Matrix multiplication & $\mathcal{O}(QN)$ \\
        Fast Fourier Transform & $\mathcal{O}(iN\log iN)$ \\
        Quadratic Interpolation & $\mathcal{O}(iN\log iN + Q)$\\
        Singular Value Decomp. & $\mathcal{O}((K_R + K_I)(N + Q))$ \\
        \hline
    \end{tabular}
    \label{tab:svd_summary}
\end{table}

\section{RESULTS}

We perform experiments to analyze the accuracy of the previously described GCC-PHAT methods: Matrix Multiplication (MM), Fast Fourier Transform with interpolation rate from 1 to 32 (FFT01, FFT02, FFT04, FFT08, FFT16 and FFT32), Fast Fourier Transform with interpolation rate from 1 to 32, followed by Quadratic Interpolation (FFT01-QI, FFT02-QI, FFT04-QI, FFT08-QI, FFT16-QI, and FFT32-QI), and Singular Value Decomposition (SVD).

Table \ref{tab:results_parameters} lists the parameters used with the investigated methods.
The search space modeled by an $180^{\circ}$ arc is discretized by 181 points, which provides a resolution of one degree.
The STFT is computed with frames of $32$ msecs, spaced by intervals of $10$ msecs.
Microphones are spaced by $d = 5$ cm, which matches typical distances on smart speakers.
The speed of sound $c$ is chosen to match normal indoor conditions, and the reconstruction error $\delta$ is set to a small value to make the SVD method accurate.

\begin{table}[!ht]
    \caption{GCC-PHAT Parameters}
    \vspace{2pt}
    \centering
    \def\arraystretch{1.0}
    \begin{tabular}{|ccccccc|}
    \hline
        $Q$ & $N$ & $\Delta N$ & $d$ & $c$ & $f_S$ & $\delta$ \\
    \hline
        $181$ & $512$ & $160$ & $0.05$ & $343.0$ & $16000$ & $10^{-5}$ \\
    \hline
    \end{tabular}
    \label{tab:results_parameters}
\end{table}

Simulations are conducted to measure the accuracy of the proposed method.
The two microphones are positioned randomly in rooms of various sizes.
Three categories of rooms are investigated: small (between 5m x 5m x 3m and 10m x 10m x 5m), medium (between 10m x 10m x 3m and 20m x 20m x 5m) and large (between 20m x 20m x 5m and 20m x 20m x 10m).
For each configuration, the room category is chosen randomly, and then the dimensions are generated randomly within the corresponding size range.
The reflection coefficients are set to 0, 0.3 and 0.6 to investigate different reverberation levels.
Reverberation is modeled with Room Impulse Responses (RIRs) generated with the image method \cite{allen1979image}.
Sound segments from the TIMIT dataset are then convolved with the generated RIRs \cite{zue1990speech}.
Diffuse white noise is added to each channel, for a signal-to-noise ratio (SNR) that varies from $10$dB to $40$dB.
A total of 1000 different configurations are generated.

The Root Mean Square Error (RMSE) is calculated in Eq.~\ref{eq:results_rmse} by summing all DOAs for each room and speech signal, with the estimated DOA weighted with the energy $E_{est}$, which provides insight on the relevance of each DOA.
\begin{equation}
    RMSE = \left\lVert (\sum\theta_{est}E_{est})/(\sum E_{est}) - \theta_0\right\rVert_2
    \label{eq:results_rmse}
\end{equation}
where $\lVert\dots\rVert_2$ stands for the Euclidean norm.

\begin{figure*}[!ht]
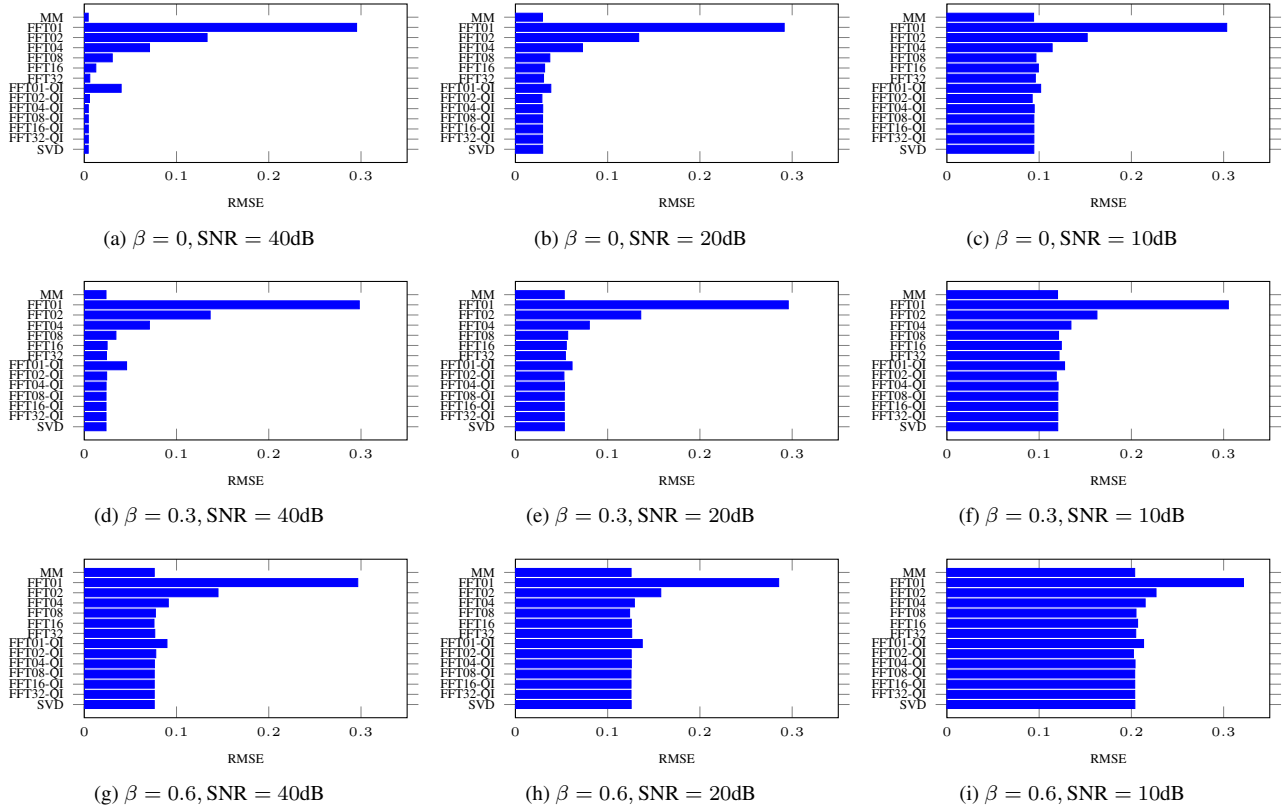

    \centering
    \subfloat[$\beta = 0, \textrm{SNR} = 40\textrm{dB}$]{\bargraphrsme{data/rsme_betaLow-snrHigh.dat}{0.33\textwidth}{105pt}}
    \subfloat[$\beta = 0, \textrm{SNR} = 20\textrm{dB}$]{\bargraphrsme{data/rsme_betaLow-snrMid.dat}{0.33\textwidth}{105pt}}
    \subfloat[$\beta = 0, \textrm{SNR} = 10\textrm{dB}$]{\bargraphrsme{data/rsme_betaLow-snrLow.dat}{0.33\textwidth}{105pt}}
    \\
    \subfloat[$\beta = 0.3, \textrm{SNR} = 40\textrm{dB}$]{\bargraphrsme{data/rsme_betaMid-snrHigh.dat}{0.33\textwidth}{105pt}}
    \subfloat[$\beta = 0.3, \textrm{SNR} = 20\textrm{dB}$]{\bargraphrsme{data/rsme_betaMid-snrMid.dat}{0.33\textwidth}{105pt}}
    \subfloat[$\beta = 0.3, \textrm{SNR} = 10\textrm{dB}$]{\bargraphrsme{data/rsme_betaMid-snrLow.dat}{0.33\textwidth}{105pt}}
    \\
    \subfloat[$\beta = 0.6, \textrm{SNR} = 40\textrm{dB}$]{\bargraphrsme{data/rsme_betaHigh-snrHigh.dat}{0.33\textwidth}{105pt}}
    \subfloat[$\beta = 0.6, \textrm{SNR} = 20\textrm{dB}$]{\bargraphrsme{data/rsme_betaHigh-snrMid.dat}{0.33\textwidth}{105pt}}
    \subfloat[$\beta = 0.6, \textrm{SNR} = 10\textrm{dB}$]{\bargraphrsme{data/rsme_betaHigh-snrLow.dat}{0.33\textwidth}{105pt}}
    \caption{RMSE for the investigated methods with different reflection coefficients ($\beta$) and SNR -- smaller is better.}
    \label{fig:results_rmse}
\end{figure*}

Figure \ref{fig:results_rmse} shows the RMSE with multiple reverberation levels and SNR values.
Note how the accuracy decreases as the reverberation increases (when the reflection coefficient $\beta$ increases) and as the SNR decreases.
For all cases, the MM method provides the best performance.
Moreover, the FFT\{02,04,08,16,32\}-QI matches the MM performance, followed closely by the SVD approach.
It is interesting to note the reduction in accuracy when the quadratic interpolation alone on non-interpolated FFT results (FFT01-QI) is used.
Note how the FFT interpolation also reduces the RMSE, but requires a significant factor (FFT32) to achieve a performance level similar to MM, whereas other interpolation rates increase the RMSE (FFT\{01,02,04,08,16\}).

\begin{figure}[!ht]
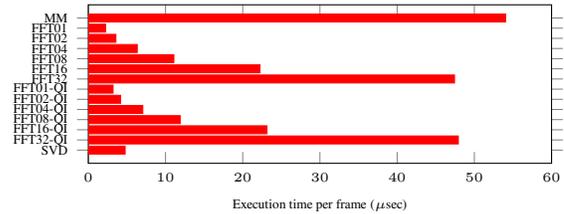

    \centering
    \bargraphcpu{data/cpu.dat}{0.9\columnwidth}{105pt}
    \caption{Execution time per frame on a PC -- less is better.}
    \label{fig:results_cpu}
\end{figure}

All methods are implemented in C to minimize execution time on an Intel Xeon E5-1620 3.70GHz (source code is available online\footnote{http://github.com/FrancoisGrondin/gccphat}), and use the FFTW library to optimize the computations of FFTs \cite{frigo1998fftw}.
Figure \ref{fig:results_cpu} shows the execution time per frame for all methods.
Among methods that minimize RMSE, FFT02-QI and SVD also minimize the execution time.
As expected, the MM approach is the most expensive in terms of computational requirements, and the FFT interpolation increases rapidly the execution time.
Note that, as opposed to the FFT from the FFTW library, the matrix multiplication in the SVD method is performed elementwise, and not optimized for special sets of instruction such as Streaming SIMD Extensions (SSE) \cite{raman2000implementing}.
Using these special instructions should reduce the SVD method execution time.

\section{CONCLUSION}

In this paper we investigate some GCC-PHAT interpolated-based methods and a SVD-based method for accurate DOA estimation for physically close microphones.
Results demonstrate that interpolation by increasing the size of the FFT and quadratic interpolation must be combined together to perform accurate DOA estimation.
It also demonstrates that the proposed SVD transform also offers good accuracy.
Finally, doubling the size of the FFT with quadratic interpolation and the SVD method both offer the best trade-off between accuracy and computational load.
As future work, we could evaluate the performance for various distances between a pair of microphones, and then choose the optimal GCC-PHAT building blocks with microphone arrays of arbitrary shapes.

\vfill\clearpage

\bibliographystyle{IEEEbib}
\bibliography{refs}

\end{document}